\documentclass[draft, onecolumn]{IEEEtran} 

\usepackage{amsmath, amssymb,  graphicx,  algorithm}
\usepackage{subcaption}
\usepackage{bbm,dsfont,nicefrac,mathrsfs}
\usepackage{mathtools,algorithmic}
\usepackage{times}

\newlength\myindent
\newcommand\bindent{%
	\begingroup
	\setlength{\itemindent}{\myindent}
	\addtolength{\algorithmicindent}{\myindent}
}
\newcommand\eindent{\endgroup}

\title{Blind Exploration and Exploitation of Stochastic Experts}

\author{Noyan~C.~Sev\"uktekin~and~Andrew~C.~Singer\\Department of Electrical and Computer Engineering\\
	University of Illinois at Urbana-Champaign\\
	sevukte2@illinois.edu}%

\newcommand\given{\;\middle|\;}
\newcommand{\nn}{\nonumber}
\newcommand{\lb}{\left\{}
\newcommand{\rb}{\right\}}
\newcommand{\myset}[1]{\lb#1\rb}
\newcommand{\brac}[1]{\left[#1\right]}
\newcommand{\abs}[1]{\left|#1\right|}
\newcommand{\yay}[1]{\left(#1\right)}
\newcommand{\ind}[1]{\mathds{1}\left(#1\right)}

\newcommand{\prob}[1]{\mathbb{P}\left(#1\right)}
\newcommand{\condprob}[2]{\mathbb{P}\left(#1\given #2 \right)}
\newcommand{\expt}[1]{\mathbb{E}\left[#1\right]}
\newcommand{\V}[1]{V\left(#1\right)}

\newcommand{\condexpt}[2]{\mathbb{E}\left[#1\given#2\right]}
\newcommand{\pp}[1]{\tilde{p}_{#1}}

\newcommand{\comm}[1]{\mathcal{C}_{#1}}
\newtheorem{lemma}{Lemma}

\newtheorem{prop}{Proposition}

\DeclareMathOperator*{\argmin}{arg\,min}
\DeclareMathOperator*{\argmax}{arg\,max}
\DeclareMathOperator{\sign}{sign}

\begin{document}

\maketitle

\begin{abstract}
	We present blind exploration and exploitation (BEE) algorithms for identifying the most reliable stochastic expert based on formulations that employ posterior sampling, upper-confidence bounds, empirical Kullback-Leibler divergence, and minmax methods for the stochastic multi-armed bandit problem. Joint sampling and consultation of experts whose opinions depend on the hidden and random state of the world becomes challenging in the unsupervised, or blind, framework as feedback from the true state is not available. We propose an empirically realizable measure of expert competence that can be inferred instantaneously using only the opinions of other experts. This measure preserves the ordering of true competences and thus enables joint sampling and consultation of stochastic experts based on their opinions on dynamically changing tasks. Statistics derived from the proposed measure is instantaneously available allowing both blind exploration-exploitation and unsupervised opinion aggregation. We discuss how the lack of supervision affects the asymptotic regret of BEE architectures that rely on UCB1, KL-UCB, MOSS, IMED, and Thompson sampling. We demonstrate the performance of different BEE algorithms empirically and compare them to their standard, or supervised, counterparts. 
\end{abstract}


\section{Introduction}
The standard stochastic multi-armed bandit framework captures the exploration-exploitation trade-off in sequential decision making problems under partial feedback constraints. The objective is to actively identify the best member, or members, of a community comprising stochastic sources, termed arms, while suffering the relative loss of non-ideal choices. Often, the arms yield rewards that belong to a known probability distribution with hidden parameters, and upon choosing an arm, the decision maker observes the reward from that arm directly. For ergodic reward distributions, \cite{gittins1979bandit} shows that the  dynamic programming solution takes the form of an index policy\footnote{The orignial formulation of \cite{gittins1979bandit} was in the Bayesian framework.}, called dynamic allocation indices, which motivated the rich body of work that led to the arm-selection rules that eventually achieved the asymptotic regret lower bounds of \cite{lai1985asymptotically}. Alternatively, when the probability law that governs the rewards is defined conditionally with respect to a hidden state that represents the ``changing world'', restless bandit framework of \cite{whittle1988restless} leads to arm-selection policies that are often computationally demanding. A key challenge that we address here is to develop index policies that identify the best source in an environment where the underlying state of the world changes erratically and hence, the observations are generated from different probability distributions at each point in time.  

Specifically, we consider the case where each arm represents a stochastic expert providing opinions on changing tasks and thus, upon consulting an expert, the decision maker observes an opinion, rather than a direct reward. Stochastic experts are sources of subjective information that might fail but not purposefully deceive, as discussed in \cite{cesa2006prediction}, and often, expert suggestions, or opinions, are used with the aid of side-information: Feedback from past states of the world is used in boosting, \cite{schapire2012boosting}, models of expert stochasticity, or direct information of expert reliability, or competence, are often used in the Bayesian framework, \cite{poor2013introduction}. In the absence of \textit{any} side information, the decision maker operates in a regime that can be termed \textit{unsupervised}, relying solely on the information in the opinions. Unsupervised opinion aggregation methods such as expectation maximization (EM) \cite{welinder2010online}, belief propagation (BP) \cite{karger2011iterative}, and spectral meta-learner (SML) \cite{parisi2014ranking} exhibit an interesting phenomenon: The reliability of experts are inferred as side-product of the underlying optimization for estimating past states based on a block of opinions. On the other hand, joint sampling and consultation of experts without supervision, or blind exploration and exploitation (BEE) as termed here, requires instantaneously available statistics that would allow reliable inference of expert reliabilities at any and all states of the world. 

We propose a method that relies solely on opinions to infer the competence of an expert by re-defining the notion competence as the probability of agreeing with peers rather than being objectively correct. The proposed method does not only allow empirical inference of competence without any supervision but also enables the use of index policies to efficiently address exploration and exploitation dilemma when the underlying task changes at random. We show that standard, or supervised, exploration-exploitation (SEE) strategies extend their uses to the BEE problem by consulting multiple experts for each task, equivalent to sampling multiple arms in the standard framework. Specifically, we consider the index rules that rely on posterior sampling, \cite{thompson1933likelihood}, upper-confidence bounds such as UCB1, \cite{auer2002finite}, and KL-UCB, \cite{garivier2011klucb}, minimum empirical Kullback-Leibler divergence, in particular, IMED, \cite{honda2015imed}, and minmax rule MOSS of \cite{audibert2009minimax}. We investigate two operational regimes: First,
a fixed number of experts are consulted for each task and the opinion of the expert who is believed to be most-reliable at that time is chosen. Second, upon consulting a group of experts, a decision is formed by aggregating their opinions without further supervision. We empirically compare the performance of different BEE index rules and demonstrate that exploration-exploitation-based choice of experts leads to comparable results to those of the original algorithms in the unsupervised framework. 

The organization of this paper is as follows: We summarize the notation used in this paper, provide a background on stochastic experts, and define the BEE problem formally in Section \ref{sec:probdef}. We discuss the motivation, formal definition, and properties of our technique for unsupervised reliability inference in Section \ref{sec:pseudocomp}. Then, we discuss the fundamental properties of the BEE index rules in Section \ref{sec:bee}. The experiments for comparing different BEE algorithms as well as comparing them to their SEE counterparts are in Section \ref{sec:experiments}. The proofs are deferred to the appendix.   

\section{Notation, Background, and Problem Formulation}
\label{sec:probdef}
We begin with a brief overview of the notation used in this paper. Then, we formally define the key concepts regarding stochastic experts. We conclude this section by defining the BEE problem.    
\subsection{Notation}
A probability space is a triplet $\yay{\Omega, \mathscr{F}, \mathbb{P}}$, where $\Omega$ is the event space, $\mathscr{F}$ is the sigma-field defined on $\Omega$, and $\mathbb{P}$ is the probability measure. Random variables are denoted by capital letters with the corresponding samples being denoted by lowercase letters: $(X,x)$. A random process is an indexed collection of random variables: $\myset{Y(t): t\in \mathbb{T}}$, where $\mathbb{T}$ is the index set. Independent random variables $\yay{X_1, X_2}$ are denoted by $X_1 \perp X_2$ and conditionally independent random variables $\yay{X_1, X_2}$ conditioned on $Y$ are denoted by $X_1 - Y - X_2$. Expectation, conditional expectation, and conditional probability operators are denoted by $\expt{\cdot}$, $\condexpt{\cdot}{\cdot}$, and $\condprob{\cdot}{\cdot}$ respectively. The indicator function is denoted by $\ind{\cdot}$, where domain is to be understood from context. We use  $[T] \triangleq \myset{1,\cdots,T}$ to denote the positive natural numbers up to a finite limit $T<\infty$. All logarithms $\yay{\log}$ are taken with respect to the natural base. We use big $O$ notation when necessary. 
\subsection{Background}
Conceptually, stochastic experts are honest-but-fallible computational entities that do not deceive the decision maker deliberately. Here, we consider experts that do not collaborate while generating their opinions; \cite{cesa2006prediction} provides a detailed discussion. The goal of this paper is to propose techniques that identify the best stochastic experts, while dynamically consulting others on varying tasks. In that context, consulting an expert on a task is equivalent to pulling an arm in the standard multi-armed bandit framework. The true reward, however, remains hidden.     

Formally, let us begin with a random process $\myset{Y(t): t\in [T]}$ that represents binary states of the world, or tasks with binary labels: $Y(t) \in \myset{-1,1}$, $\forall t\in[T]$. We allow the nature to generate tasks independently:
\begin{equation}
	\label{independent_tasks}
	Y\yay{t_1} \perp Y\yay{t_2},~\forall t_1\neq t_2 \in [T].
\end{equation}
Furthermore, let the random process $Y(t)$ that governs the evolution of tasks  maximize the uncertainty: 
\begin{equation}
	\label{unif_distr}
	\prob{Y(t) =1} = \prob{Y(t)=-1} = \nicefrac{1}{2},~\forall t\in [T].
\end{equation} 
It is worth noting that any bias from non-uniform task generation can either be estimated directly from labeled data, or inferred without supervision via methods such as \cite{jaffe2016unsupervised}. Furthermore, while independence assumption appears to be restrictive, it is common in stochastic multi-armed bandit formulations, \cite{bubeck2012regret}.   

Formal characterization of stochastic experts involves the reliability of their opinions and statistical dependence to the others. The probability with which the opinion of an expert identifies the true state of the world correctly determines the reliability, or competence, of that expert: 
\begin{equation}
	\label{static_competence}
	p_i \triangleq \prob{X_i(t) =Y(t)},~\forall t\in[T].
\end{equation} 
Here, the reliability of an expert does not depend on the underlying state of the world \footnote{A notable exception to this model is the ``two-coin'' model from \cite{dawid1979maximum}, where conditionally static competences are discussed.}. 
We further allow that experts generate opinions $\myset{X_i(t): i\in [M]}$ independently from one another for every task $t\in[T]$. Formally: 
\begin{equation}
	\label{independent_generation}
	X_i(t) - Y(t) - X_j(t),~\forall i\neq j\in [M],~\forall t\in[T].
\end{equation} 
Conceptually, it makes sense that for meaningful inference, two different opinions on the same task should never be statistically independent. Furthermore, experts having conditionally independent opinions is equivalent to independence of rewards in the standard framework.

Given the probability law defined by eq.~\eqref{independent_tasks}-\eqref{independent_generation}, we can formally discuss why SEE algorithms requires a toolset to address the impact of the underlying uncertainty. Observe that:
\begin{equation}
	\label{motivation}
	\lim\limits_{t\rightarrow \infty} \frac{1}{t} \sum_{\tau =1}^{t} X_i(t) = 0,~\forall p_i\in\brac{0,1},
\end{equation} 
which follows from the law of total probability, see appendix \ref{app:average_opinion}. Conceptually, eq.~\eqref{motivation} indicates that the average opinion does not reflect the competence of an expert, which is the true reward, posing a challenge for joint exploration and exploitation in the context of sequentially consulting stochastic experts, which we formally define next.  

\subsection{Problem Definition}
The first objective of the BEE problem is to identify the best expert in a population while actively consulting members of that group on tasks that change from one consultation to another. The following notion of regret, written here in normalized form, formally captures this phenomenon:  
\begin{equation}
	\label{real_regret}
	R_T =  \frac{1}{T}\sum_{t=1}^{T} \ind{X^{*}(t)=Y(t)} - \frac{1}{T}\sum_{t=1}^{T} \ind{X_{I_t}(t)=Y(t)}.
\end{equation}
Here $X^{*}$ is the opinion of the most competent expert; $X^{*} = X_{i^{*}}$, where $i^{*} = \argmax_{i\in[M]} p_i$ and $I_t\in[M]$, $\forall t\in[T]$ is the expert chosen at time $t$. Observe that the regret, as defined in eq.~\eqref{real_regret} depends on the sample path of opinions and hence, it is difficult to analyze rigorously. Nonetheless, it simplifies asymptotically:
\begin{equation}
	\label{real_regret_asymptotic}
	\lim\limits_{T\rightarrow \infty}R_T = \max_{i\in[M]} p_i - \lim\limits_{T\rightarrow \infty}\frac{1}{T}\sum_{t=1}^{T} \ind{X_{I_t}(t)=Y(t)}.
\end{equation}
The first term is a direct consequence of the ergodicity of the processs $\ind{X^{*}(t)=Y(t)}$, which follows directly from eq.~\eqref{independent_tasks}-\eqref{static_competence}. Conceptually, this amounts to the fact that one can measure the true reliability of an expert given sufficiently many labeled tasks, as long as the reliability of the expert does not change across tasks, as is the case here. 

Motivated by similar asymptotic behaviors, a notion of \textit{pseudo regret} often arises in the context of stochastic bandits, see, for instance, \cite{bubeck2012regret}. In the context of stochastic experts, the pseudo regret is defined as follows:  
\begin{equation}
	\label{pseudo_regret}
	\tilde{R}_T = \max_{i\in[M]} p_i - \frac{1}{T}\expt{\sum_{t=1}^{T} \ind{X_{I_t}(t)=Y(t)}}.
\end{equation}
Another notion of pseudo regret provides a reliable metric for the performance of BEE rules that aggregate opinions after consulting experts. Let a $m<M$ experts be consulted for every task $t$, indexed by $\comm{t}\subset[M]$ and let $f: \myset{\pm 1}^{\abs{\comm{}}}\rightarrow \myset{\pm 1}$ be a known opinion aggregation rule. Define the pseudo regret as: 
\begin{equation}
	\label{swarm_regret}
	\tilde{R}_T = \max_{\substack{\comm{}\subset[M]\\ \abs{\comm{}}=m}} \prob{f\yay{\myset{X_i(t):i\in\comm{}}}=Y(t)} - \frac{1}{T}\expt{\sum_{t=1}^{T} \ind{f\yay{\myset{X_i(t):i\in\comm{t}}}=Y(t)}}.
\end{equation}
To emphasize the difference in objectives, we will refer to the rules that aim to minimize eq. \eqref{pseudo_regret} as BEE rules and to those that sequentially weight and aggregate opinions of reliable members (SWARM) and thus, aim to minimize eq. \eqref{swarm_regret}, as SWARM rules.   

Given an arm-selection strategy, it is more intuitive to structure a theoretical analysis based on the pseudo regrets in eq.~\eqref{pseudo_regret}-\eqref{swarm_regret}, provided the true rewards $\myset{\ind{X_{I_{\tau}}(\tau))=Y(\tau)} : \tau \in [t]}$ are known. Nonetheless, the main challenge remains: At any given time $t$, the decision maker does not have access to the past labels $\myset{Y_{\tau}: \tau \in [t]}$ and thus, does not get to observe the true rewards. We address this challenge by consulting multiple experts for each task and leveraging the increased reliability of a collection of experts for inferring those of the individuals. Next, we formally define and discuss the properties of the competence estimation technique that facilitates joint exploration and exploitation based on opinions alone.    
%
%

\section{Unsupervised Estimation of Competences with Instantaneous Updates}
\label{sec:pseudocomp}
Two seemingly conflicting statistical phenomena should be reconciled to facilitate sampling and consultation of experts based only on their opinions. On one hand, standard index policies for  exploration and exploitation require observations from a joint probability distribution that does not evolve in time. On the other hand, opinions $\myset{X_1(t),\cdots,X_M(t)}$ are based on the hidden state $Y(t)$ and thus, their statistics change randomly in time. Furthermore, an operational challenge exists: Multi-armed bandit problems often allow observations to be available instantly so that the player can decide which arm to pull next. This is in sharp contrast to the unsupervised opinion aggregation strategies such as SML, EM, BP from \cite{parisi2014ranking,welinder2010online,karger2011iterative} respectively, that yield the reliability of experts as a side-product of an optimization process that is often carried out over a block of opinions.

Fortunately, there exists an empirically realizable measure of competence that is only based on opinions. Conceptually, consider a compact measure, for instance, an ideal ruler for measuring the length of a two-dimensional line. The length of any such line can be measured reliably by this ruler and these measurements can be used to determine which line is the longest. Recall that the true competence of an expert can be measured experimentally via:
\begin{equation}
	\lim\limits_{t\rightarrow \infty} \frac{1}{t} \sum_{\tau=1}^{t} \ind{X_i(t)=Y(t)} = p_i. 
\end{equation}      
In other words, state feedback is an ideal ruler in our example. In the absence of such reliable reference, one can \textit{choose to accept} a random line as a measure of length, replacing the phrase ``this line is $\ell$ units long'' with ``this line is $\beta$ \textit{measuring lines} long''. This strategy cannot, of course, infer the true length of any line however, it can infer the \textit{relative length} of every other line, which is sufficient for constructing index rules that solve the BEE problem.     

We introduce an alternative notion of reliability, termed \textit{pseudo competence}, that allows instantaneous, opinion-based inference of the competences, while preserving the true ordering of the competences. Formally, the pseudo competence of an expert is defined as follows:
\begin{equation}
	\label{pseudocompdef}
	\pp{i} = \prob{X_i(t) = \V{\myset{X_j(t): j\in\comm{}}}} = \prob{X_i(t) = \sign\brac{\sum_{j\in\mathcal{C}}X_j(t)}}.
\end{equation} 
Here, $\V{\cdot}$ denotes majority voting (where ties are broken arbitrarily\footnote{The $\yay{\sign}$ operator does not conventionally ``break ties arbitrarily''. However, we allow $\sign\brac{0}=Z$, where $\prob{Z=1}=\prob{Z=-1}=\nicefrac{1}{2}$ in eq.~\eqref{pseudocompdef} for notational clarity.}) over a committee $\mathcal{C}$ that excludes the $i^{th}$ expert: $\mathcal{C}\subset [T]\setminus\myset{i}$. Observe that it is possible to estimate pseudo competences empirically: 
\begin{equation}
	\label{empirical_estimate}
	\lim\limits_{t\rightarrow \infty} \frac{1}{t} \sum_{\tau=1}^{t} \ind{X_i(t)=\V{\myset{X_j(t): j\in\comm{}}}} = \pp{i}. 
\end{equation}   
Further note that similar to the standard formulation with direct rewards, one can easily keep track of number of times an expert is consulted and the number of times that expert agreed with the others. 

Let us denote the probability of the committee $\mathcal{C}$ being correct by:
\begin{equation}
	p_{\mathcal{C}} \triangleq \prob{\V{\myset{X_j(t): j\in\comm{}}}}.
\end{equation}
Then, the pseudo competence can be written explicitly as:
\begin{equation}
	\label{pseudocomp_explicit}
	\pp{i} = p_i p_{\mathcal{C}} + \yay{1-p_i} \yay{1-p_{\mathcal{C}}}. 
\end{equation}
The proof of eq.~\eqref{pseudocomp_explicit} appears in appendix \ref{app:prop} as a part of the proof for Proposition \ref{prop}. Conceptually, the pseudo competence of an expert increases during empirical estimation if either both the expert and committee are correct or both are incorrect. 

Beyond allowing empirical estimation, the pseudo competence, as a measure of reliability, demonstrates a key property: Given pseudo competences for a set experts, or given sufficiently large number of tasks and the corresponding opinions thanks to eq.~\eqref{empirical_estimate}, it is possible to rank experts based on their true competence, as discussed next.
\begin{prop}
	\label{prop}
	For a committee $\mathcal{C}\subset [T]$, the following holds for every $i,j\notin \mathcal{C}$: 
	\begin{equation}
		\label{pushtogether}
		\pp{i}-\pp{j} = \yay{2p_{\mathcal{C}}-1}\yay{p_i-p_j}.
	\end{equation}
	Therefore, if the probability of the committee $\mathcal{C}$ making a correct decision under majority rule satisfies:
	\begin{equation}
		\label{condition}
		p_{\mathcal{C}} = \prob{\V{\myset{X_j(t): j\in\comm{}}}=Y(t)}> \nicefrac{1}{2},	
	\end{equation}
	then pseudo competence preserves the ordering of true competences: 
	\begin{equation}
		p_i\geq p_j \iff \pp{i} \geq \pp{j}.
	\end{equation}
\end{prop}
The proof is given in appendix \ref{app:prop}. It is worth noting that while the condition in eq.~\eqref{condition} is not too restrictive, it is also unavoidable. Conceptually, this condition amounts to the committee $\mathcal{C}$ of peers being collectively reliable. Equivalently, it is not possible to infer the reliability of an expert based on opinions from a group that often collectively fails to identify the true state of the world. Formally, we restrict our attention to the case where experts are reasonably reliable: $p_i>\nicefrac{1}{2}$, $\forall i\in[M]$, which implies that eq.~\eqref{condition} holds for every subset $\mathcal{C}\subset[M]$ of experts. We further discuss the definition and the properties of the pseudo competence for the experts that belong to the committee $i,j\in\mathcal{C}$ in appendix \ref{app:morepseudo}.


\section{Fundamental Properties of Different BEE and SWARM Rules}
\label{sec:bee}
There is a rich literature governing how to jointly sample and consult experts when feedback from the true state of the world $\myset{Y(\tau): \tau\in[t-1]}$ is available to the player. In this section, we propose unsupervised applications of different exploration-exploitation strategies using pseudo competence as reward. Specifically, we focus on UCB1, \cite{auer2002finite}, due to its use of the Chernoff-Hoeffding bound, KL-UCB, \cite{garivier2011klucb}, due its optimality for the Bernoulli rewards, IMED, \cite{honda2015imed}, due to its near-optimal performance and empirical robustness, MOSS, \cite{audibert2009minimax}, due to its minmax optimality, and Thompson sampling due to the strong synergy between posterior sampling and pseudo competences. The proposed BEE algorithms have the following form:

\begin{algorithm}
	\caption{BEE Algorithm}
	\begin{algorithmic}
		\STATE \textbf{Require:} Choice of $\Psi_i(\cdot)$ for exploration-exploitation strategy, as defined in eq. \eqref{ucb1stat}-\eqref{thompson_stat}.
		\STATE \textbf{Initialize:} 
		\bindent
		\STATE Consult every expert on the first task, 
		\STATE Update agreements via eq. \eqref{unsupervised_reward}, estimate competences via \eqref{empirical_pseudo} with $\comm{1}=[M]$.
		\eindent
		\STATE \textbf{Loop:}  
		\bindent
		\STATE $\comm{t}\leftarrow$ Pick top $m$ (bottom for IMED) experts ranked by $\Psi_i(t)$,
		\STATE $I_t \leftarrow \argmax_{i\in[\comm{t}]} \Psi_i(t)$ ($\argmin$ for IMED),
		\STATE Commit to $X_{I_t}$ as the player decision,\\
		\STATE Update agreements via eq. \eqref{unsupervised_reward}, estimate competences via \eqref{empirical_pseudo} with $\comm{t}$.
		\eindent
	\end{algorithmic}
\end{algorithm}

As discussed in Section \ref{sec:pseudocomp}, the notion of pseudo competence is built upon the availability of peer opinions therefore, we will consider consulting multiple experts for each task $t\in[T]$, equivalent to playing multiple arms, see, for instance, \cite{komiyama2015optimal}. Let $\comm{t}$ denote the set of experts consulted for task $t\in[T]$ and define the reward from consulting an expert as whether that expert agrees with the collective decision of the peers for that task: 
\begin{equation}
	\label{unsupervised_reward}
	R_i(t;\comm{t}) = \ind{X_i(t) = \V{\myset{X_j\yay{t}:~j\in\comm{t}\setminus\myset{i}}}}.
\end{equation} 
Observe that contrary to the state feedback, opinions $\myset{X_j:j\in\comm{t}}$ are available from any subset $\comm{t}\subset [M]$ of experts therefore, the empirical pseudo competence estimate for an expert at time $t$ takes the form:  
\begin{equation}
	\label{empirical_pseudo}
	\pp{i,T_i(t-1)} \triangleq  \frac{1}{T_{i}(t-1)}\sum_{
		\tau\in[t-1]:~i\in \comm{\tau}} R_i(\tau;\comm{\tau}).
\end{equation}	
We will now use the statistics on agreements among experts as defined in eq. \eqref{unsupervised_reward} and the concomitant empirical competence estimation in eq. \eqref{empirical_pseudo} to form the relevant statistics $\Psi_i(t)$ for sampling and consulting experts in the unsupervised setup.  

To begin with the upper-confidence bound strategies, UCB1 leads to consulting experts that have highest ranking in:
\begin{equation}
	\label{ucb1stat}
	\Psi_i^{\text{UCB1}}(t) = \pp{i,T_i(t-1)}+\sqrt{\frac{2\log t}{T_i(t-1)}}.
\end{equation}
Particularly in the Bernoulli case, several refinements to UCB1 algorithm exist: Notably, \cite{garivier2011klucb} shows that Kullback-Leibler upper confidence bound (KL-UCB) algorithm achieves the regret lower bound of \cite{lai1985asymptotically}. In the unsupervised setup, experts with the highest ranking according to the following (operational)\footnote{\cite{garivier2011klucb} remark that the threshold $\log\yay{t}$ should be taken as $\log\yay{t}+c\log\log\yay{t}$ for analytical purposes, where they recommend $c=0$ in practice. Furthermore, we experimentally implement its more robust counterpart KL-UCB+, see \cite{honda2019note}.} statistics are consulted:
\begin{equation}
	\label{klstat}
	\Psi_i^{\text{KL-UCB}}(t) = \max\yay{q\in\brac{0,1}:~ T_i(t-1)~ d\yay{\pp{i,T_i(t-1)}, q}\leq \log\yay{\frac{t}{T_i\yay{t-1}}}},
\end{equation} 
where Kullback-Leibler divergence $d\yay{\cdot,\cdot}$ is defined as:
\begin{equation*}
	d\yay{p,q} = p\log\frac{p}{q} + \yay{1-p}\log\frac{1-p}{1-q}.
\end{equation*}
Albeit more computationally demanding, the performance of the KL-UCB algorithm makes it an appealing choice for binary opinions. Alternatively, using indexed minimum empirical divergence (IMED) algorithm of \cite{honda2015imed}, we consult the experts that have the lowest ranking according to:
\begin{equation}
	\label{imed_stat}
	\Psi_i^{\text{IMED}}(t) = T_i(t-1)~d\yay{\pp{i,T_i(t-1)}, \max_{i\in[M]}\pp{i,T_i(t-1)}} + \log T_i(t-1).
\end{equation}
Another key idea is that of minmax optimality, which is achieved by MOSS in the supervised framework and can be applied to the unsupervised framework by selecting experts with highest ranking in: 
\begin{equation}
	\label{moss_stat}
	\Psi_i^{\text{MOSS}}(t) = \pp{i,T_i(t-1)} + \sqrt{\frac{\max\yay{\log\yay{\frac{T}{M T_i\yay{t-1}}} , 0}}{T_i\yay{t-1}}}.
\end{equation}

Finally, Thompson sampling (TS) suggests randomly sampling arms based on the posterior distribution over competences. Specifically, 
\begin{equation}
	\label{thompson_stat}
	\Psi^{\text{TS}}_{i}\yay{t} = Z_i, \text{ where } Z_i \sim \text{Beta}\yay{\alpha_i(t),\beta_i(t)}.
\end{equation} 
The prior is often taken be uniform for every expert, $\yay{\alpha_i(0),\beta_i(0)} = \yay{1,1}$, $\forall i\in[M]$ and, in the Bernoulli case, update rules of the beta distribution have a well-defined form, see, for instance, \cite{russo2017tutorial}: 
\begin{equation*}
	\yay{\alpha_{i}(t), \beta_{i}(t)} = \yay{\alpha_{i}(t), \beta_{i}}(t) + \yay{R_{i}\yay{t,\comm{t}} + 1-R_{i}\yay{t,\comm{t}}}, \forall i\in\comm{t},
\end{equation*}
where $R_i\yay{t,\comm{t}}$ is defined in eq. \eqref{unsupervised_reward} and the statistics  $\yay{\yay{\alpha_{i},\beta_{i}}: j\in[M]\setminus \comm{t}}$ of the non-consulted experts are not updated at that time. 

As a random process, $R_i\yay{t,\comm{t}}$, as shown in eq. \eqref{unsupervised_reward}, is defined conditionally with respect to the choice of experts through $\Psi_i\yay{\cdot}$, $\forall i\in[M]$, and the resulting opinions, up to time $t$ of consulting. Therefore, it is challenging to form a complete theoretical analysis of how choosing $R_i\yay{t,\comm{t}}$ changes the the difference between the ``rewards'' in the unsupervised setup compared to those that correspond to the true competences. Nevertheless, for a fixed committee $\comm{}$ of peers, choosing $R_i\yay{t,\comm{}}$ pushes the competences together in a way that is as quantified through eq. \eqref{pushtogether} of Proposition \ref{prop}, making it possible to observe the impact of using pseudo competences on the regret of the exploration-exploitation strategies. The following result is a corollary to Proposition \ref{prop} and \cite[Theorem 2]{audibert2009minimax} for UCB1, \cite[Theorem 1]{garivier2011klucb} for KL-UCB, \cite[Theorem 5]{honda2015imed} for IMED, \cite[Theorem 6]{audibert2009minimax} for MOSS, and \cite[Theorem 1]{agrawal2013further} for Thompson sampling.

\begin{lemma}
	\label{main_lemma}
	Let the player designate a subset of experts $\comm{}\subset[M]$ as peers prior to exploration and exploitation and consult them for every task $\comm{t}=\comm{}$, $\forall t\in [T]$. Define the difference between competences:
	\begin{equation}
		\Delta_i = \max_{j\in[M]} p_j - p_i,
	\end{equation}
	and a potential function for a given set of competences:
	\begin{equation}
		\Phi = \frac{\log T}{T}\sum_{\substack{i\in[M]\setminus\comm{}\\\Delta_i>0}} \frac{1}{\Delta_i}.
	\end{equation}
	Then, the regret of a BEE algorithm that picks $I_t = \argmax_{i\in[M]\setminus\comm{}}\Psi_i(t)$ ($\argmin$ for IMED) via statistics for UCB1, eq. \eqref{ucb1stat}, KL-UCB, eq. \eqref{klstat}, or IMED, eq. \eqref{imed_stat}, is bounded by: 
	\begin{equation}
		\label{theform}
		\tilde{R}_{T}\leq C\frac{ \Phi}{2p_{\comm{}}-1},
	\end{equation}
	where the constant $C=10$ for UCB1, $C=\nicefrac{1}{2}$ for KL-UCB and IMED. The regret of Thompson sampling, eq. \eqref{thompson_stat}, also takes the form in eq. \eqref{theform} with $C=1+\varepsilon$ but with additional term of order $O\yay{\frac{M}{\varepsilon^2}}$. Furthermore, for MOSS, eq. \eqref{moss_stat}, the regret is upper bounded by: 
	\begin{equation}
		\tilde{R}_{T}^{MOSS} \leq \frac{23M}{T}\brac{\frac{1}{2p_{\comm{}}-1}\sum_{\substack{i\in[M]\setminus\comm{}\\\Delta_i>0}} \frac{\max\yay{\log\frac{T  \yay{\yay{2p_{\comm{}}-1}\Delta_i}^{2}           }{M},1}}{\Delta_i}}.
	\end{equation}
\end{lemma} 
The proof is given in appendix \ref{app:mainlemma}. The important takeaways from Lemma \ref{main_lemma} are twofold: First, using pseudo competences introduces additional regret within a constant factor of the original regret bounds but facilitates exploration and exploitation in the unsupervised regime. Second, the factor $\nicefrac{1}{\yay{2p_{\comm{}}-1}}$ diminishes as the committee $\comm{}$ collectively becomes more competent, which is important since the proposed BEE and SWARM rules that gradually select more competent experts. 

By consulting multiple experts, a set of opinions on each task is acquired and making an unsupervised decision based on such sparsely sampled opinions is often carried out by rules such as SML, EM, or BP from \cite{parisi2014ranking,welinder2010online,karger2011iterative} respectively. However, inference of pseudo competences allow an unsupervised opinion aggregation strategy that instantaneously yield a decision:  
\begin{equation}
	\label{lpnb}
	f\yay{(X_i(t),\pp{i,T_i(t)}) : i\in\comm{t}} = \sign\yay{\sum_{i\in\comm{t}}X_i(t)\yay{\pp{i,T_i(t)}-\nicefrac{1}{2}}}.
\end{equation}   
The opinion-aggregation rule in eq. \eqref{lpnb} is a linearized variant of the na\"{i}ve Bayes decision rule, which is robust to the variance of empirical averaging in \eqref{empirical_pseudo}, as discussed in \cite{berend2015finite}. In Section \ref{sec:experiments}, we compare the BEE algorithm to their SEE counterparts and we demonstrate the performance of the SWARM rule that uses linearized na\"{i}ve Bayes decision rule with pseudo competence-based weights, comparing them to their state feedback-based counterparts. 

\begin{algorithm}[t]
	\caption{SWARM Algorithm}
	\begin{algorithmic}
		\STATE \textbf{Require:} 
		\bindent
		\STATE Choice of $\Psi_i\yay{\cdot}$ for exploration-exploitation strategy, as defined in eq. \eqref{ucb1stat}-\eqref{thompson_stat}.
		\STATE Choice of $f\yay{\cdot}$ for unsupervised opinion aggregation, as defined \eqref{lpnb}
		\eindent 	
		\STATE \textbf{Initialize:} 
		\bindent
		\STATE Consult every expert on the first task, 
		\STATE Update agreements via eq. \eqref{unsupervised_reward}, estimate competences via \eqref{empirical_pseudo} with $\comm{1}=[M]$,
		\STATE Commit to majority vote: $\V{X_i:i\in[M]}$ as the first decision. 
		\eindent
		\STATE \textbf{Loop:}  
		\bindent
		\STATE $\comm{t}\leftarrow$ Pick top $m$ (bottom for IMED) experts ranked by $\Psi_i(t)$,
		\STATE Update agreements via eq. \eqref{unsupervised_reward}, estimate competences via \eqref{empirical_pseudo} with $\comm{t}$.
		\STATE Commit to $f\yay{(X_i(t),\pp{i,T_i(t)}) : i\in\comm{t}}$ as the player decision.\\
		\eindent
	\end{algorithmic}
\end{algorithm}


\section{Experiments}
\label{sec:experiments}
In this section, we discuss empirical performance of BEE and SWARM algorithms that rely on pseudo competences for reliability inference and UCB1, KL-UCB, IMED, MOSS, or Thompson sampling for exploration and exploitation. For experiments, we chose $M=100$ experts with competences chosen uniformly at random from $\brac{0.5,0.75}$ being consulted on $T=10^5$ tasks. The reason for such a competence interval is to avoid powerful experts in a community that would render the lack of supervision obsolete, similar to the use of weak classifiers in the boosting framework, \cite{schapire2012boosting}. For BEE  experiments, we consider the true normalized regret as defined in eq. \eqref{real_regret}, where for SWARM experiments, we consider the pseudo normalized regret in eq. \eqref{swarm_regret}, allowing the opinion-aggregation rule over the best committee to have direct access to the true competences. In other words, BEE algorithms are compared against supervised rules with state feedback, where SWARM algorithms compete against rules that explore and exploit with state feedback and aggregate opinions with complete information. Figure \ref{fig:bee_overall} provides an overview of BEE algorithms and Figure \ref{fig:swarm_overall} provides one for SWARM rules. 

\subsection{BEE}
We allow subcommittee sizes $m\in\brac{2,24}$, even numbers, where $m$ is the number of experts that are consulted for each task. Even number of experts chosen to make sure peers for every expert can reach a deterministic decision. 

\begin{figure}
	\centering
	\begin{minipage}{.5\textwidth}
		\centering
		\includegraphics[width=\linewidth,draft=false]{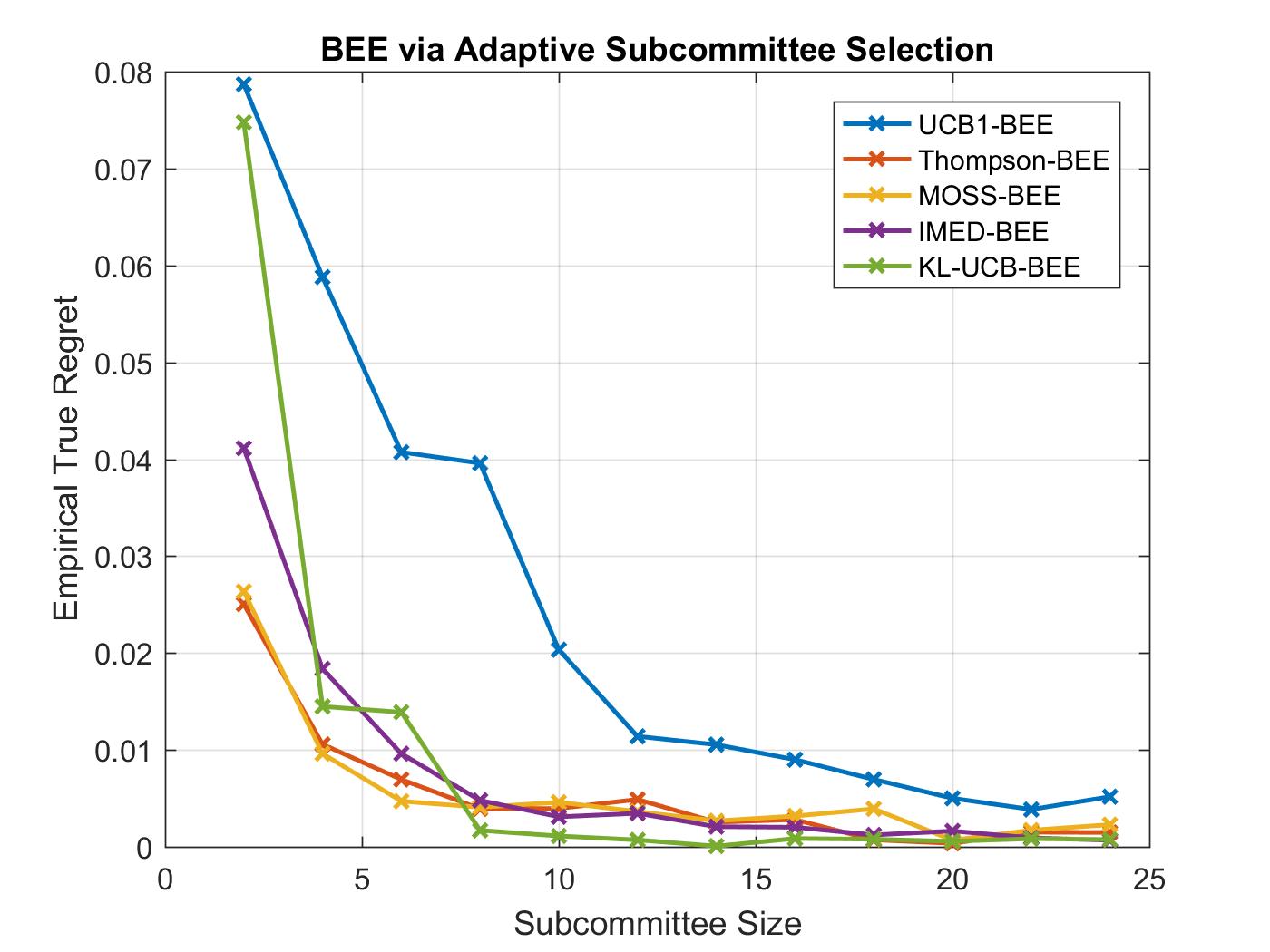}
		\caption{Performance Comparison of \\ Different BEE Algorithms}
		\label{fig:bee_overall}
	\end{minipage}%
	~
	\begin{minipage}{.5\textwidth}
		\centering
		\includegraphics[width=\linewidth,draft=false]{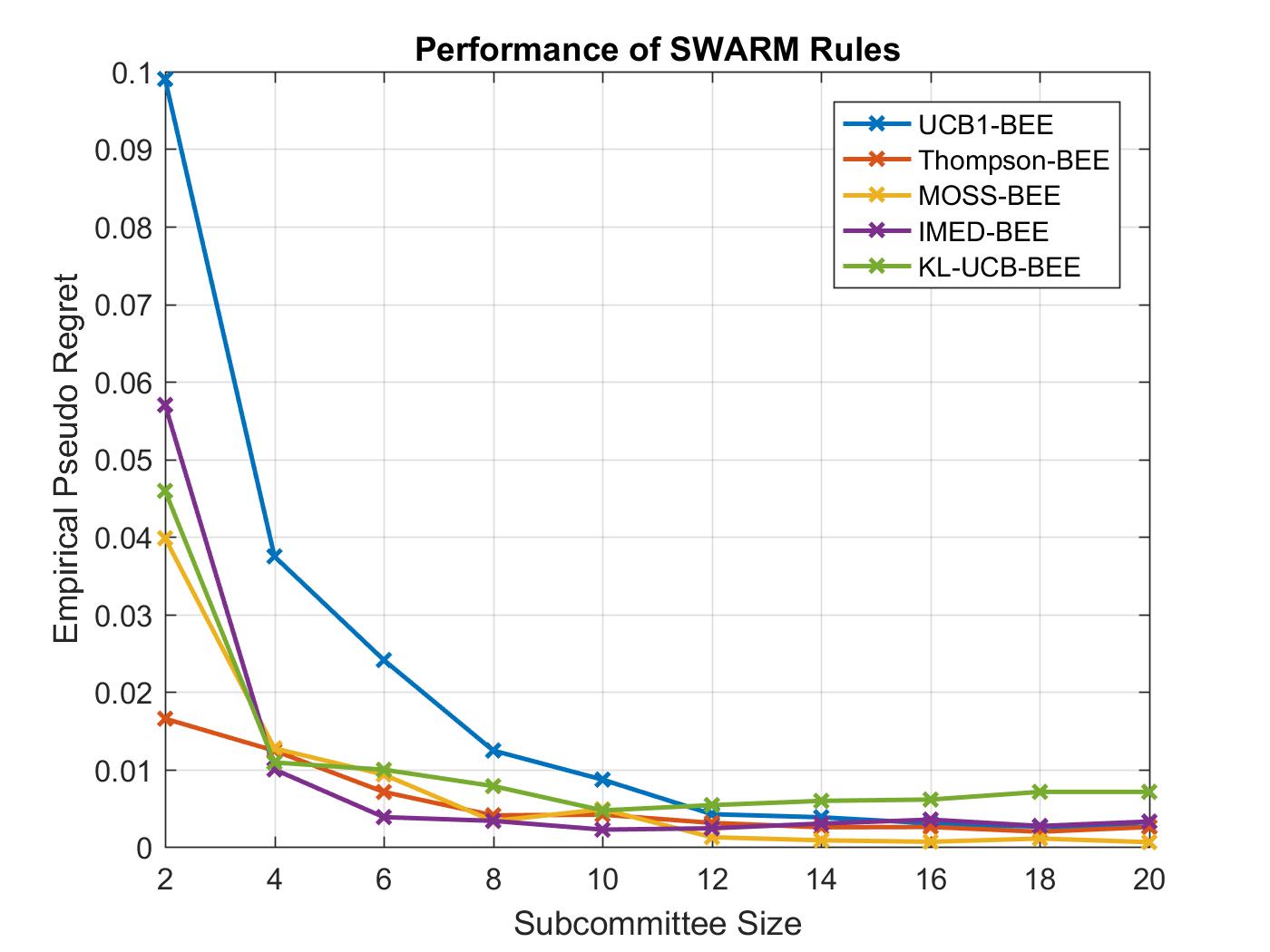}
		\caption{Performance Comparison of \\ Different SWARM Algorithms}
		\label{fig:swarm_overall}
	\end{minipage}%
\end{figure}

\begin{figure}
	\centering
	\begin{minipage}{.32\textwidth}
		\centering					
		\includegraphics[width=\linewidth,draft=false]{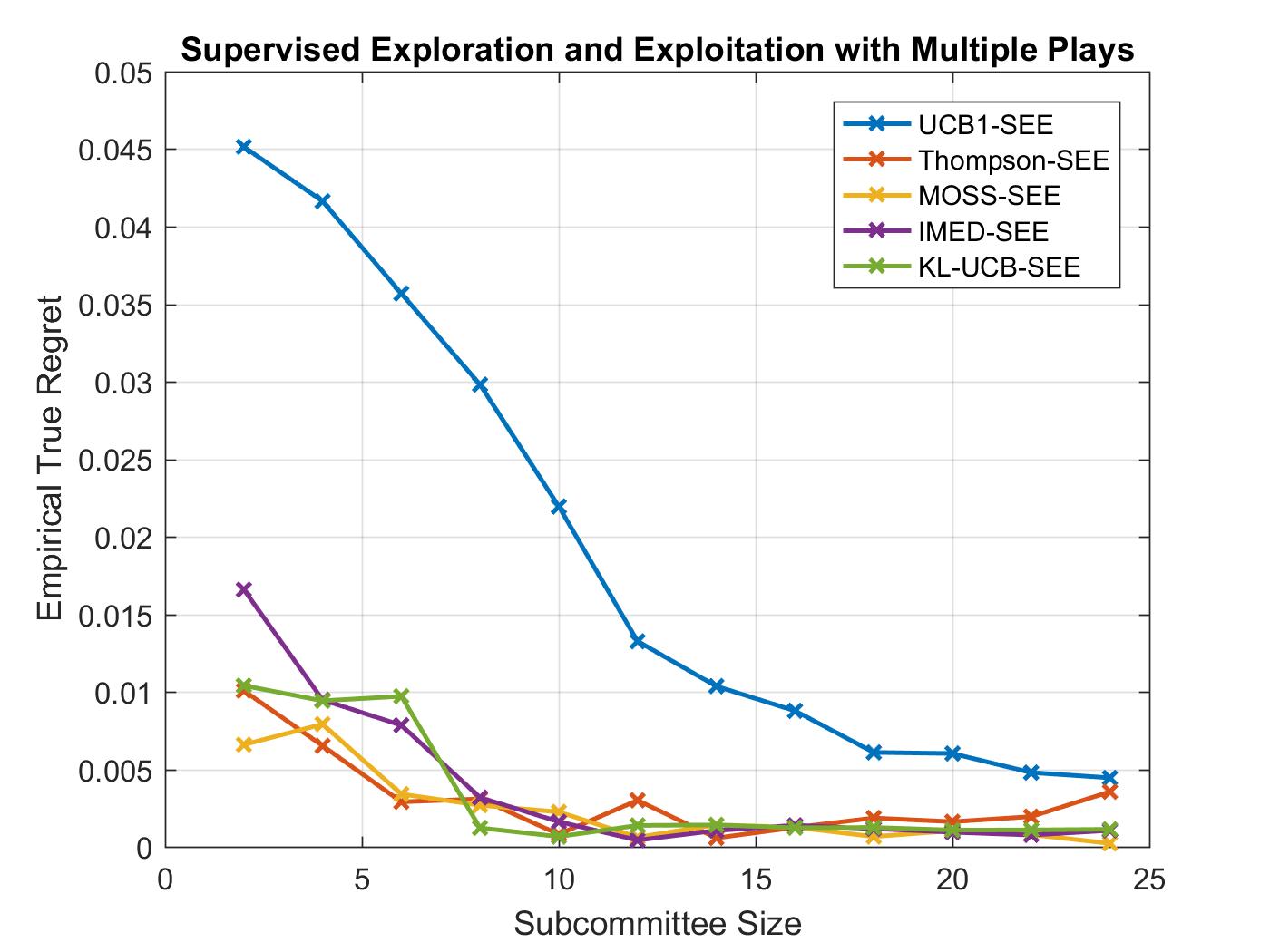}
		\caption{Overview of Multi-play SEE Algorithms}
		\label{fig:bee_reference}
	\end{minipage}
	~
	\begin{minipage}{.32\textwidth}
		\centering
		\includegraphics[width=\textwidth,draft=false]{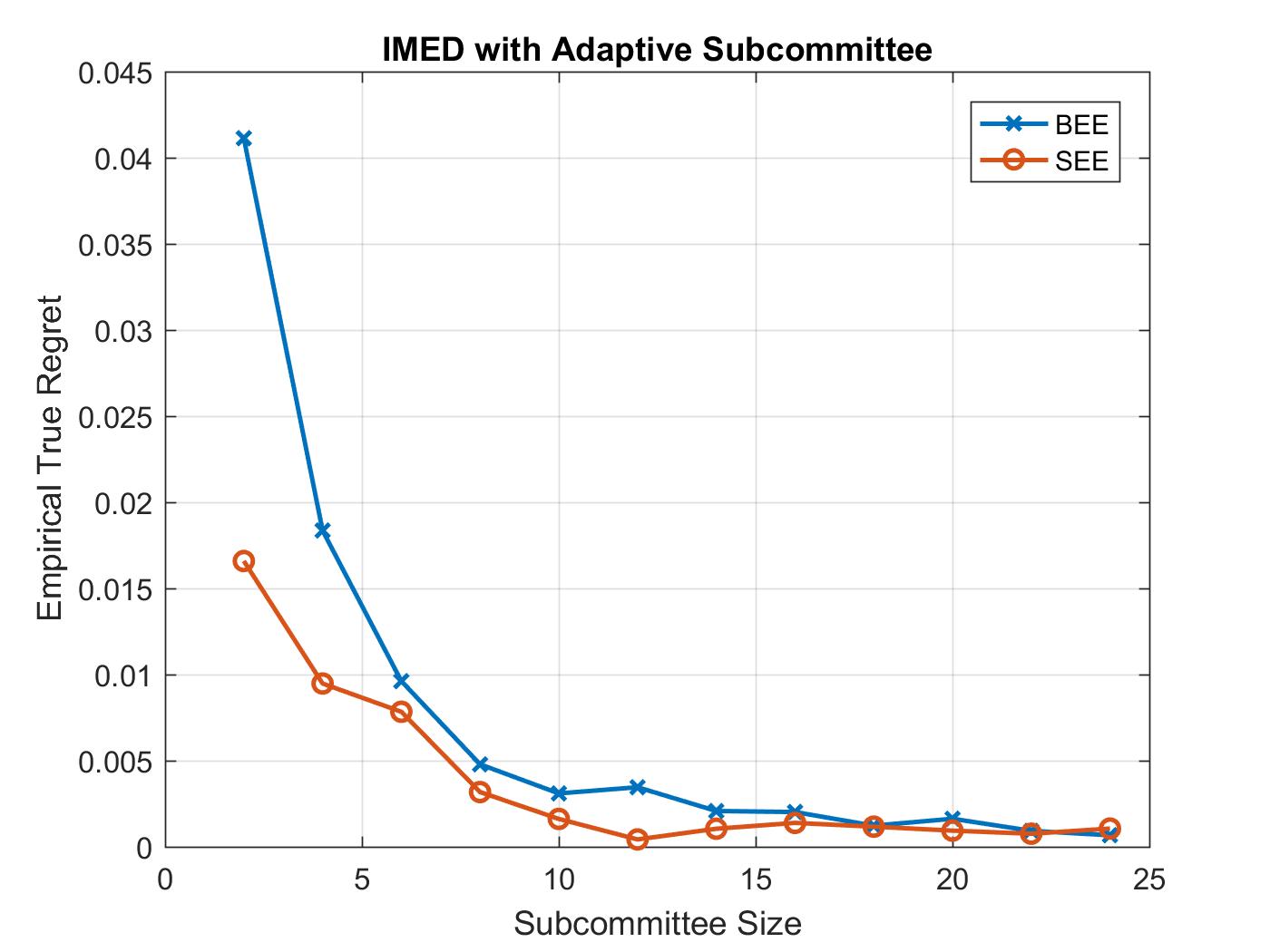}
		\caption{IMED}
		\label{fig:imedbee}
	\end{minipage}%
	~
	\begin{minipage}{.32\textwidth}
		\centering
		\includegraphics[width=\textwidth,draft=false]{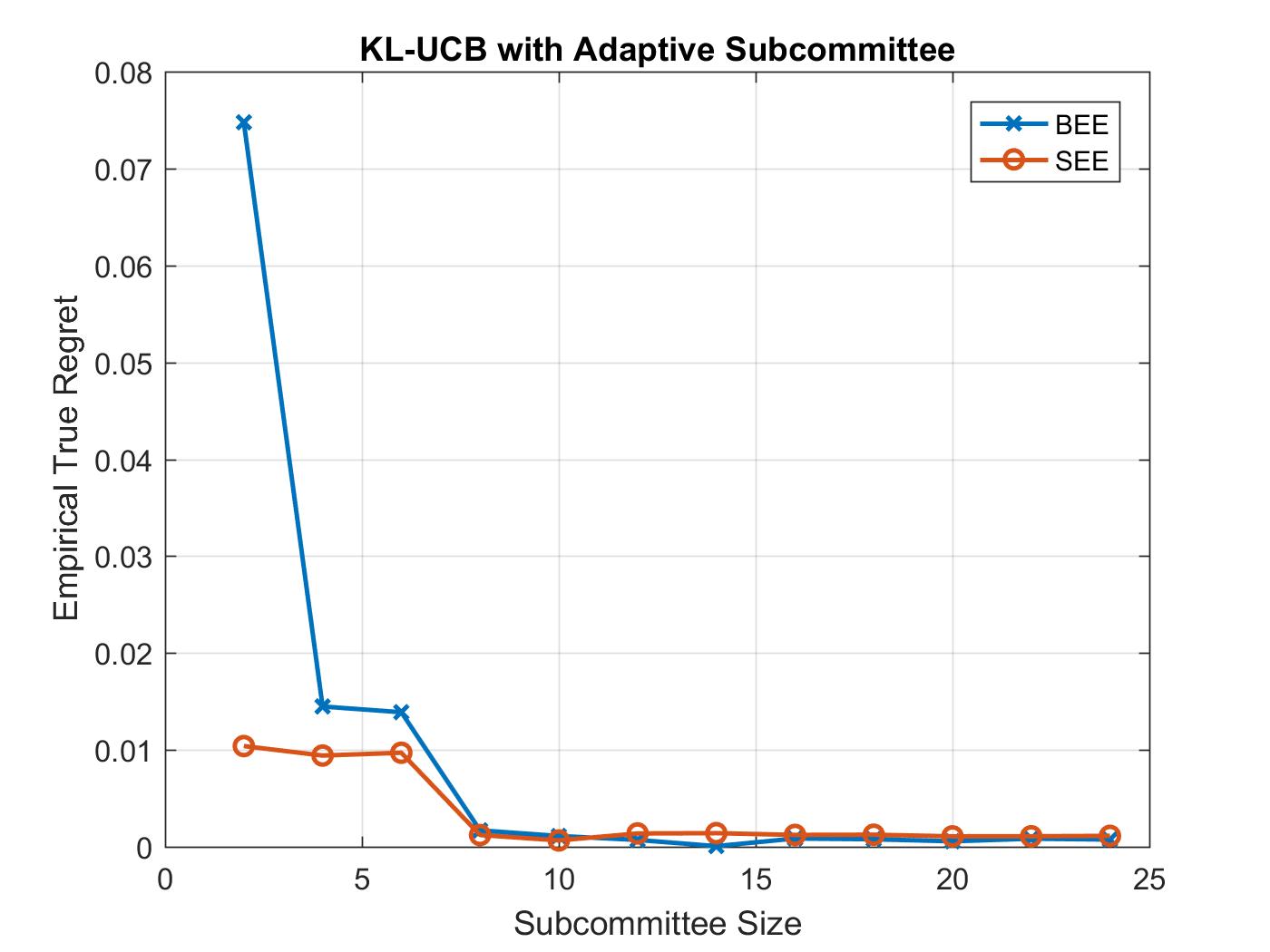}
		\caption{KL-UCB}
		\label{fig:klucbbee}
	\end{minipage}
	~
	\begin{minipage}{.32\textwidth}
		\centering
		\includegraphics[width=\textwidth,draft=false]{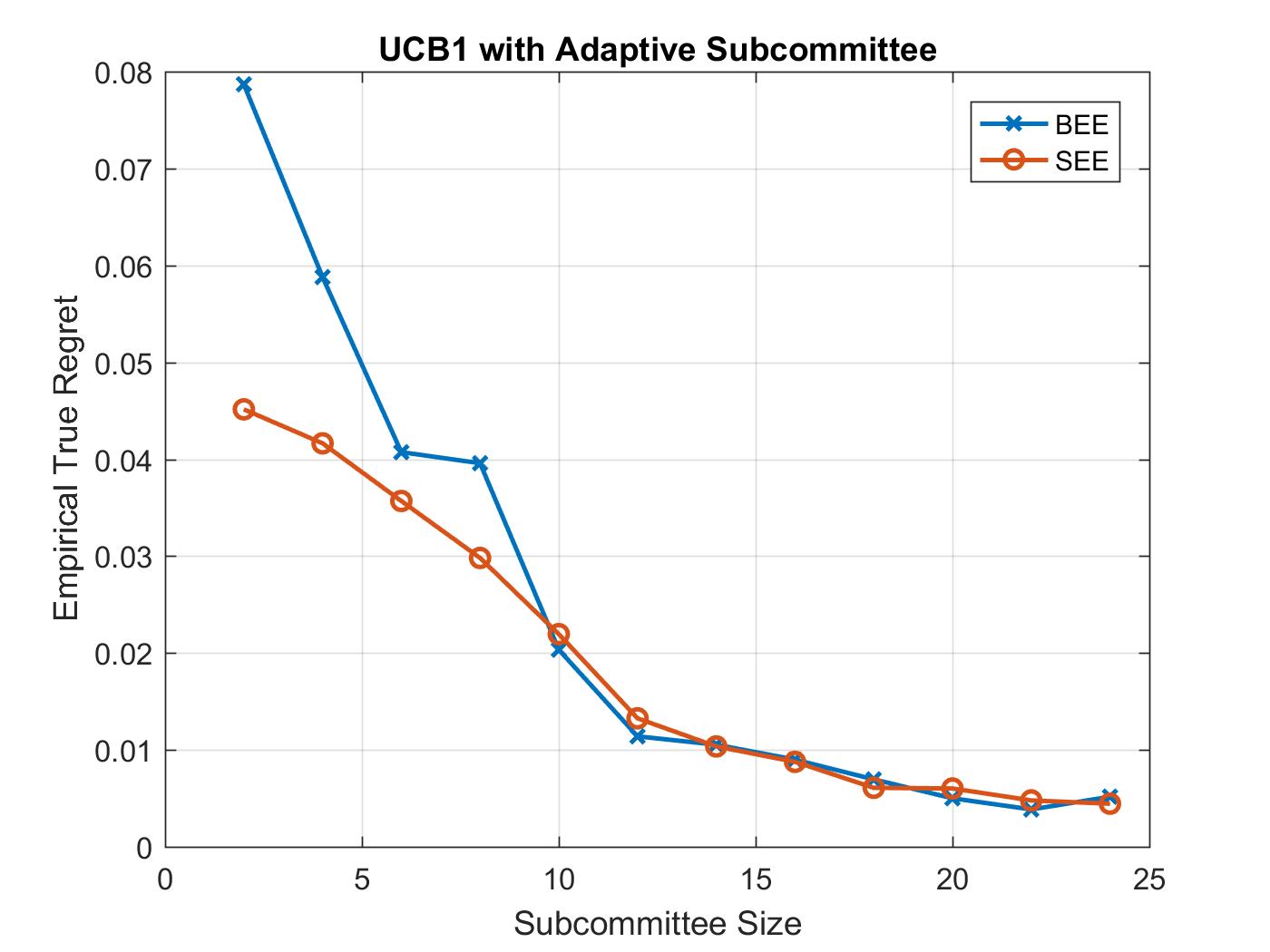}
		\caption{UCB1}
		\label{fig:ucb1bee}
	\end{minipage}%
	~
	\begin{minipage}{.32\textwidth}
		\centering
		\includegraphics[width=\textwidth,draft=false]{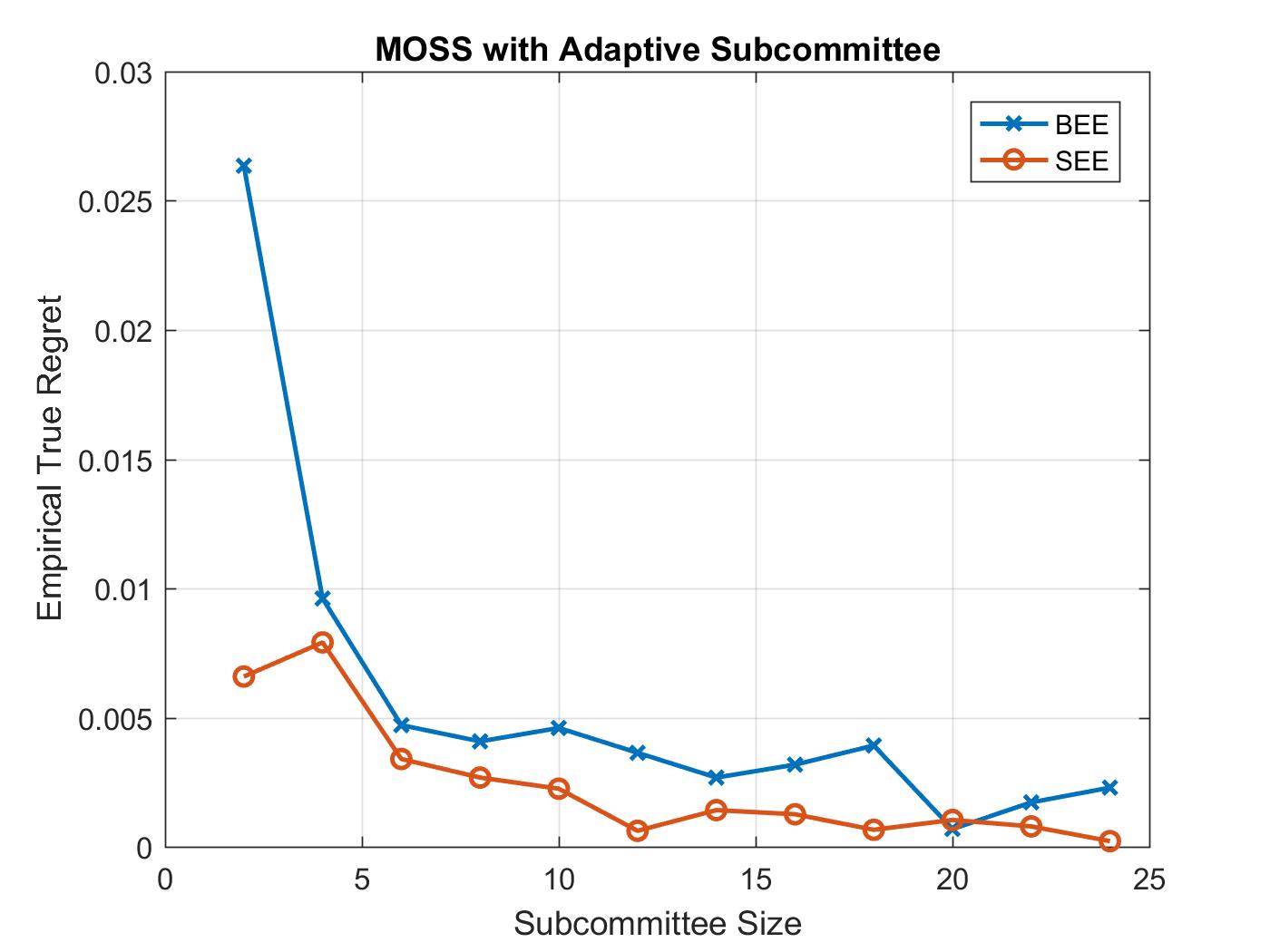}
		\caption{MOSS}
		\label{fig:mosssbee}
	\end{minipage}
	~
	\begin{minipage}{.32\textwidth}
		\centering
		\includegraphics[width=\textwidth,draft=false]{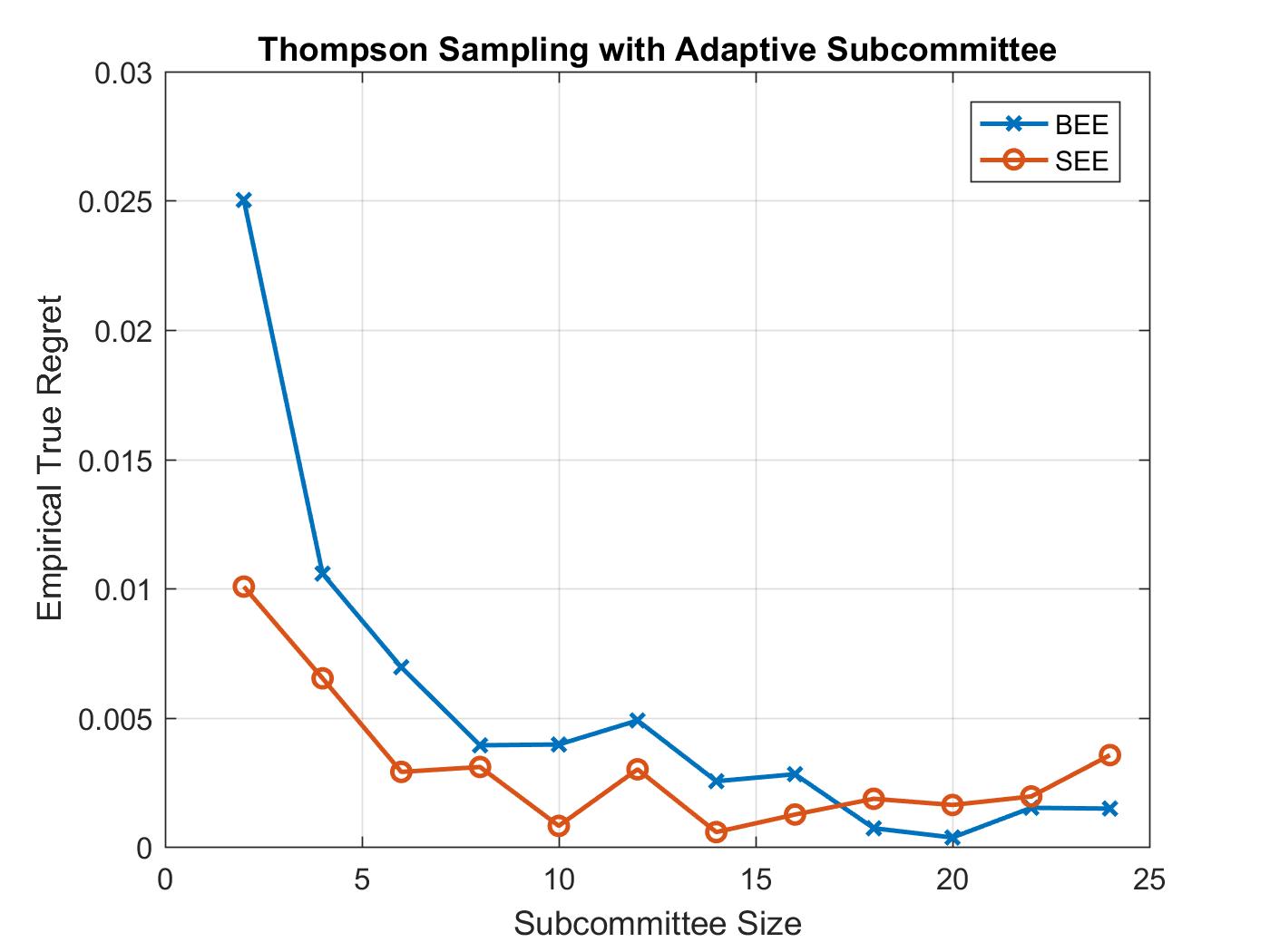}
		\caption{Thompson Sampling}
		\label{fig:thompsonbee}
	\end{minipage}
	\caption{Overview of the BEE Experiments: Performance Comparison to Supervised Counterparts}
\end{figure}

Figure \ref{fig:bee_overall} summarizes the performance of BEE algorithms that adaptively choose their subcommittees. We note that MOSS- and TS-based policies perform better than the rest for small subcommittee sizes, where KL-UCB-based BEE rule outperforms all for larger subcommittee sizes $\yay{m>8}$. Figure \ref{fig:bee_reference} illustrates the performance of UCB1, KL-UCB, Thompson sampling, MOSS, and IMED with multiple plays for reference. Interestingly, BEE algorithms that use upper-confidence bound-based exploration and exploitation strategies, as seen in Figures \ref{fig:klucbbee}-\ref{fig:ucb1bee}, perform almost identically to their SEE counterparts for large subcommittee sizes. On the other hand, the impact of supervision is clearer across the board for IMED, in Figure \ref{fig:imedbee}, MOSS, in Figure \ref{fig:mosssbee}, and Thompson sampling, as seen in Figure \ref{fig:thompsonbee}. Importantly, Figures \ref{fig:imedbee}-\ref{fig:klucbbee} indicate that BEE rules that rely on IMED and KL-UCB sequentially consult experts that on average provide opinions within $0.5\%$ of the probability of correctness of the best available expert without any supervision upon consulting modest number $\yay{m=8}$ of experts for each task.  

\subsection{SWARM}
\begin{figure}
	\centering
	\begin{minipage}{.32\textwidth}
		\centering					
		\includegraphics[width=\linewidth,draft=false]{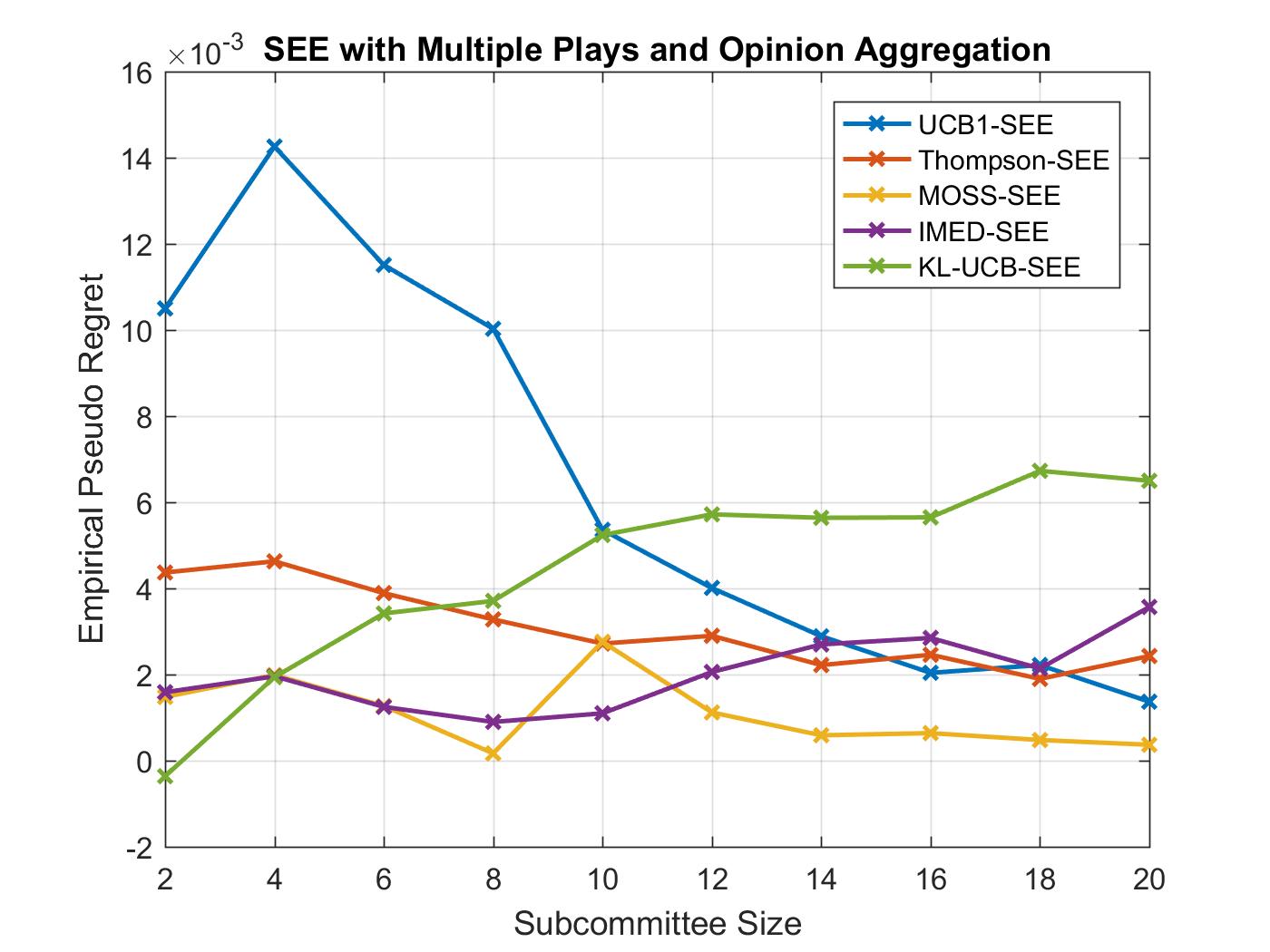}
		\subcaption{Overview Multi-play SEE \\ with Opinion Aggregation}
		\label{fig:swarm_reference}
	\end{minipage}
	~
	\begin{minipage}{.32\textwidth}
		\centering
		\includegraphics[width=\textwidth,draft=false]{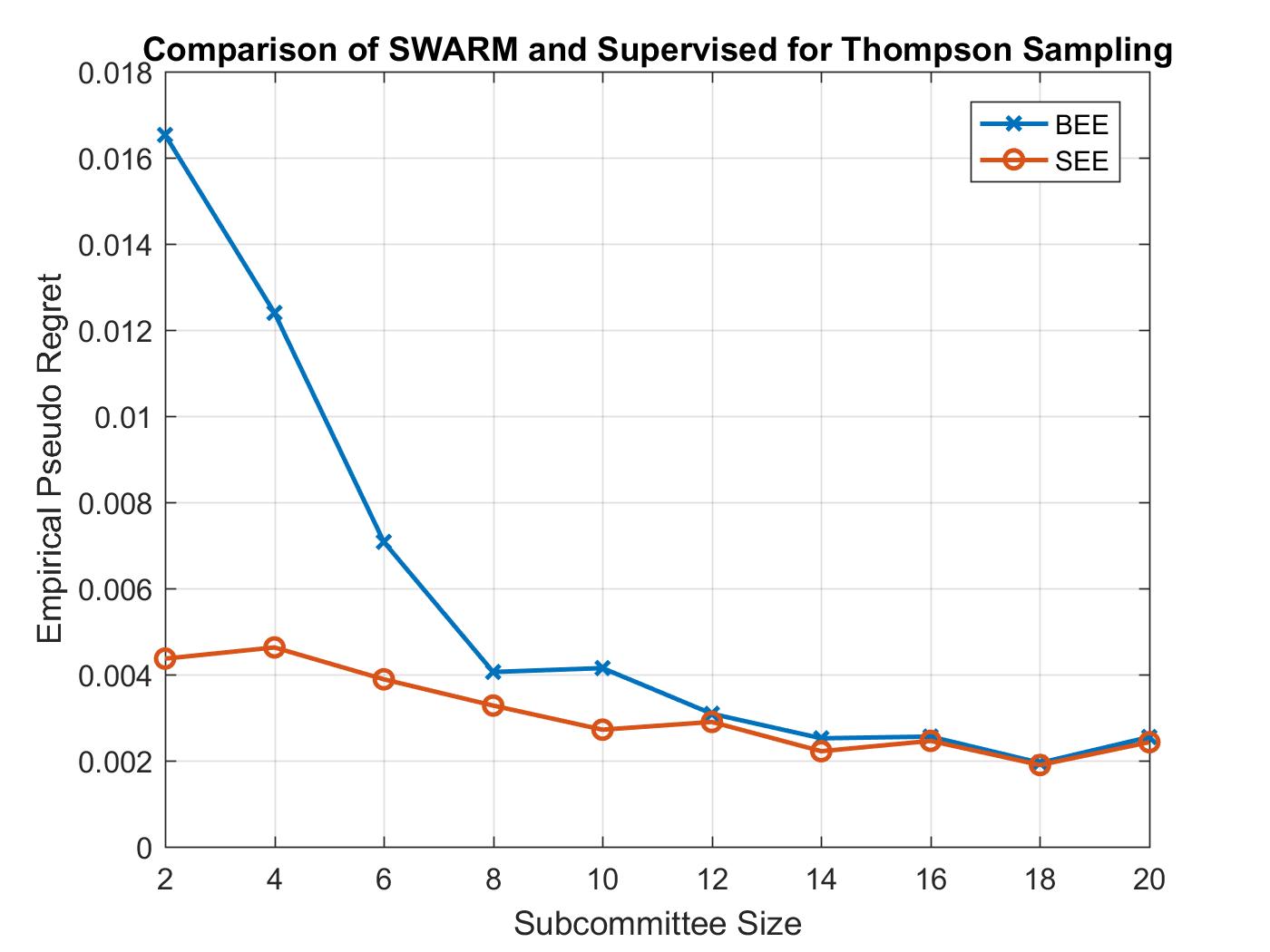}
		\subcaption{Thompson Sampling}
		\label{fig:thompsonwarm}
	\end{minipage}%
	~
	\begin{minipage}{.32\textwidth}
		\centering
		\includegraphics[width=\textwidth,draft=false]{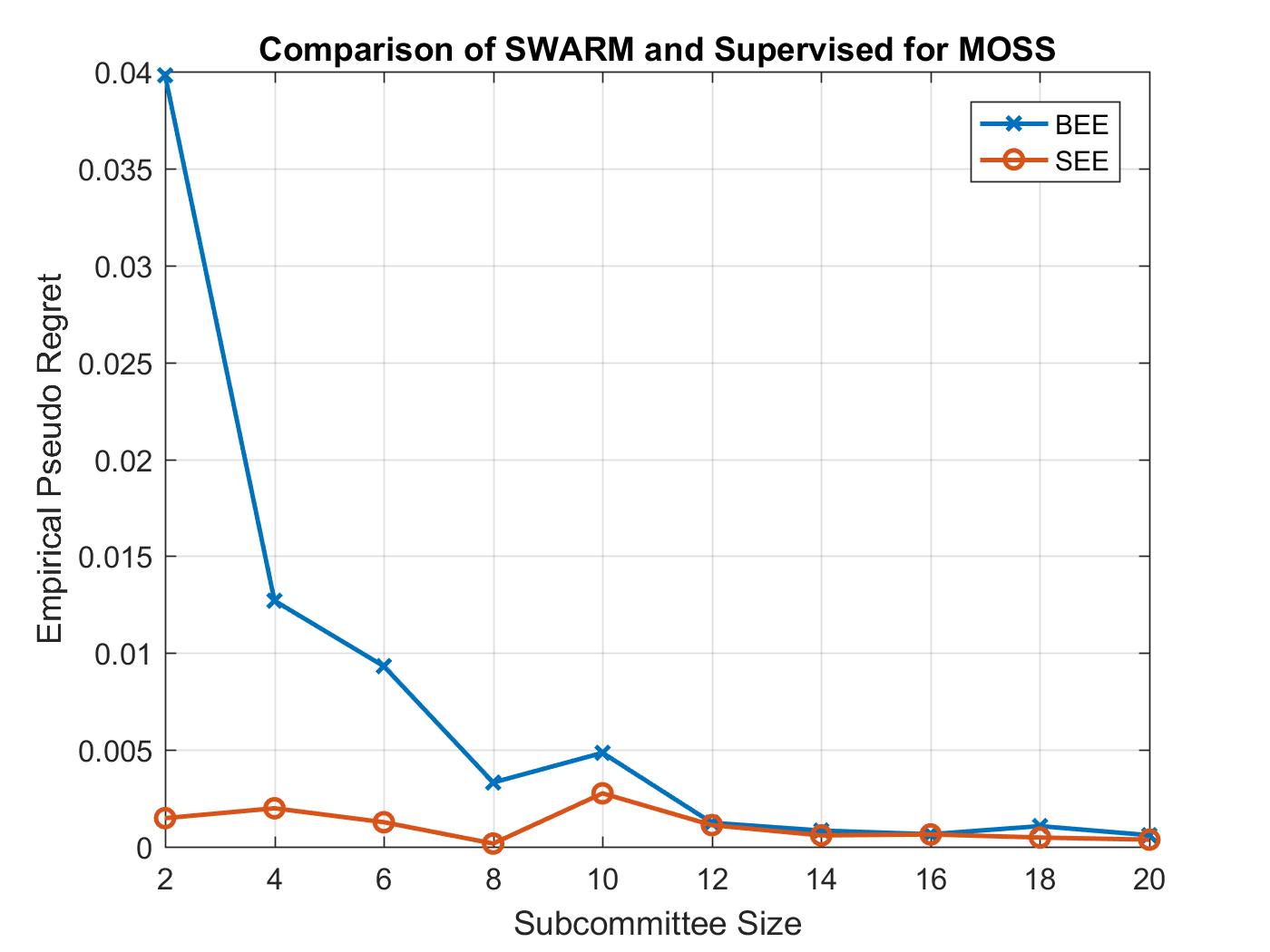}
		\subcaption{MOSS}
		\label{fig:mossswarm}
	\end{minipage}
	~
	\begin{minipage}{.32\textwidth}
		\centering
		\includegraphics[width=\textwidth,draft=false]{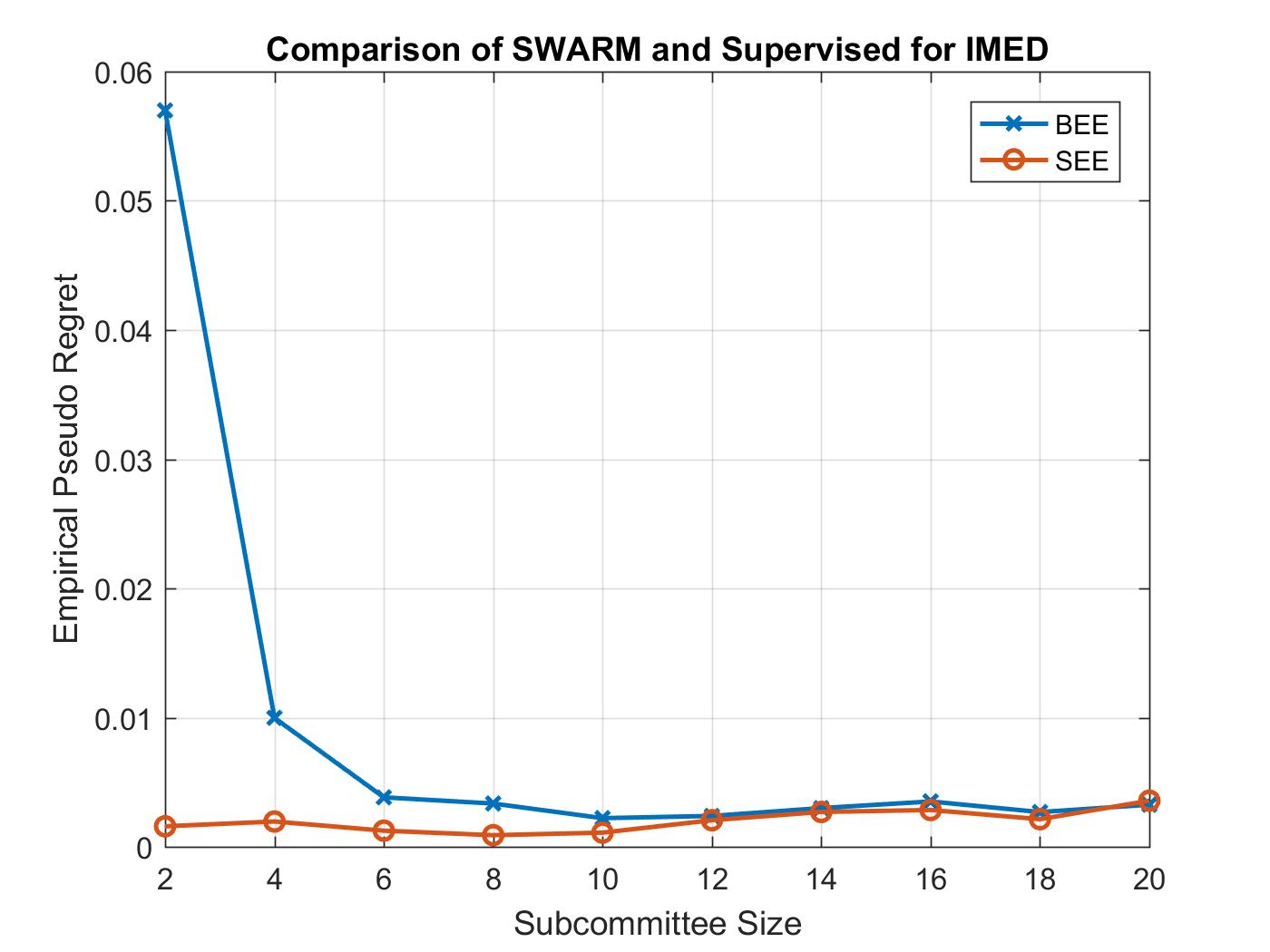}
		\subcaption{IMED}
		\label{fig:imedswarm}
	\end{minipage}
	~
	\begin{minipage}{.32\textwidth}
		\centering
		\includegraphics[width=\textwidth,draft=false]{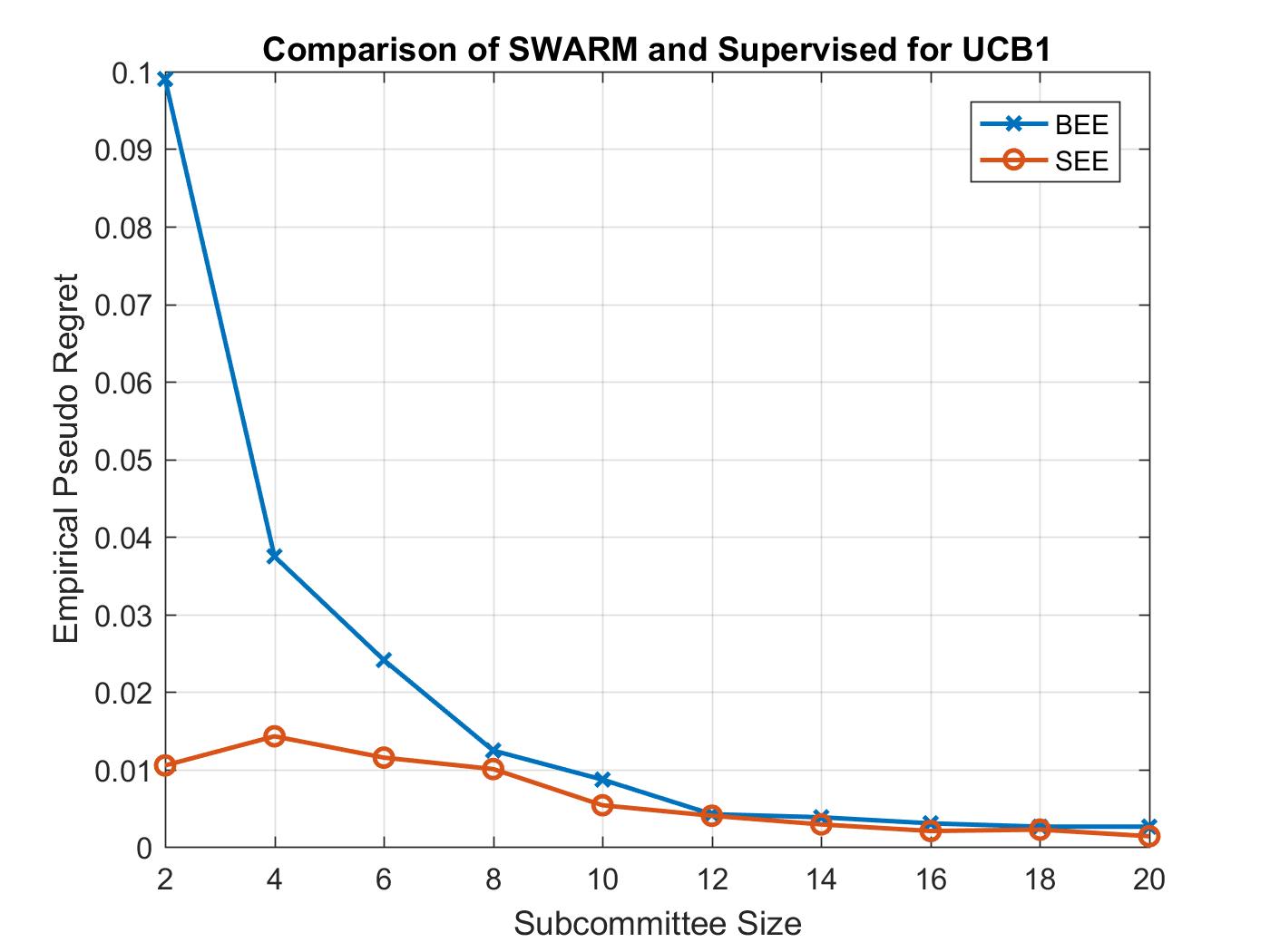}
		\subcaption{UCB1}
		\label{fig:ucb1swarm}
	\end{minipage}%
	~
	\begin{minipage}{.32\textwidth}
		\centering
		\includegraphics[width=\textwidth,draft=false]{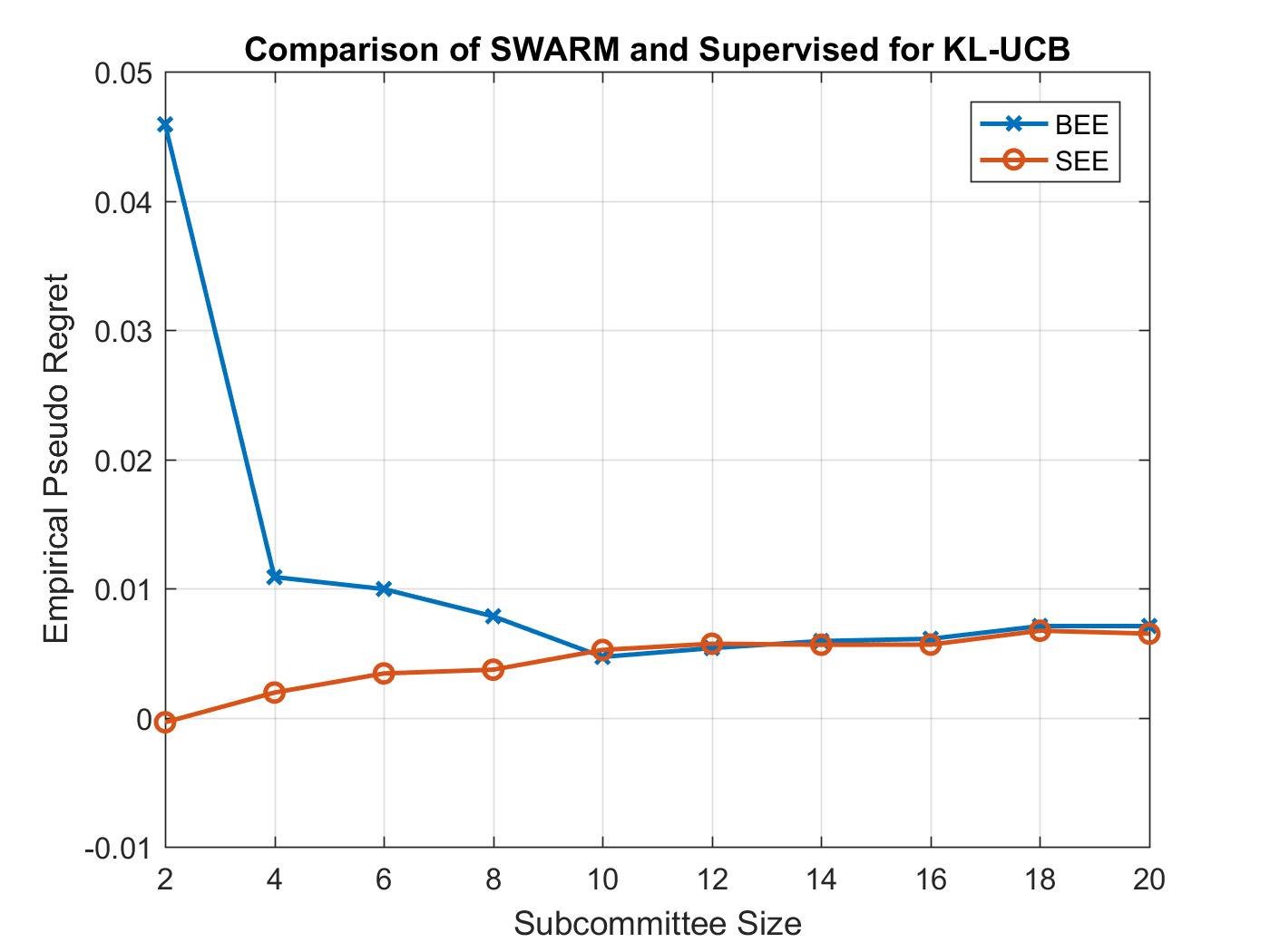}
		\subcaption{KL-UCB}
		\label{fig:klucbswarm}
	\end{minipage}
	\caption{Overview of the SWARM Experiments: Performance Comparison to Supervised Counterparts}
\end{figure}

Subcommittee sized up to $m=20$ appears to be sufficient for SWARM experiments as all of the tested rules demonstrate lower than $1\%$ normalized pseudo regret for $m>10$. Figure \ref{fig:swarm_overall} outlines the performance of the SWARM algorithms, with their supervised counterpart given in Figure \ref{fig:swarm_reference}, showing that MOSS-, IMED- and TS-based rules perform the with the additional opinion-aggregation objective. It might be expected for TS, as shown in Figure \ref{fig:thompsonwarm}, considered the synergy between consulting experts based on the posterior distribution on the competences and using an approximation of the maximum a posteriori rule that is na\"{i}ve Bayes for opinion aggregation. It is worth noting however, that the performance of IMED-based and MOSS-based consultation, which have overall robust performance in the BEE framework as shown in Figures \ref{fig:imedbee},\ref{fig:mosssbee}, stand out in the SWARM framework, as seen in Figures \ref{fig:mossswarm},\ref{fig:imedswarm}. Furthermore, Figures \ref{fig:ucb1swarm},\ref{fig:klucbswarm} indicate a similar phenomenon for the upper-confidence bound-based strategies as the BEE framework, where the performance difference between supervised and unsupervised rules diminish almost entirely. Importantly, SWARM rules achieve probability of correctness within $0.5\%$ of the opinion aggregation rule that has direct access to true competences and uses the opinions from the best available experts.  

\section{Conclusions}
In this paper, we proposed techniques that allow exploration-exploitation algorithms to operate in a regime where the rewards are observed indirectly from opinions on dynamically changing tasks. The proposed measure of competence not only allows instantaneous inference, which commonly requires either state feedback or block processing, but also enables the use of a near-optimal opinion aggregation rule in real time. Furthermore, we point out that the formal analysis of the convergence and competence ordering for BEE and SWARM algorithms are exciting open problems with the numerical evidence presented here indicating their strong potential. 
\appendix
\section{On Averaging Opinions}
\label{app:average_opinion}
Observe that the law of total probability yields the following:
\begin{equation}
	\prob{X_i(t) = 1} = \sum_{j\in\myset{\pm 1}}\condprob{X_i(t) = 1}{Y(t)=j}\prob{Y(t)=j}    
\end{equation}
Since $\prob{Y(t)=1}=\prob{Y(t)=-1}=\nicefrac{1}{2}$ and $\condprob{X_i(t)=j}{Y(t)=j}=p_i$ for every task, we can further write that:
\begin{equation}
	\prob{X_i(t) = 1} = \frac{p_i}{2} + \frac{1-p_i}{2} = \frac{1}{2},
\end{equation}
which indicates that $\expt{X_i(t)} = 0$, $\forall t$. The ergodicity of the opinion process yields eq. \eqref{motivation}. Note that this hold for every expert.
\section{Proof of Property \ref{prop}}
\label{app:prop} 
We drop the time dependence, such as $\myset{X_i(t), Y(t)}$, for notational clarity and instead write $\myset{X_i,Y}$. The proof is a direct application of the law of total probability:
\begin{align}
	\pp{i} &= \prob{X_i = \V{\myset{X_k: k\in\comm{}}}} \\  &= \prob{X_i = Y, \V{\myset{X_k: k\in\comm{}}}=Y} + \prob{X_i \neq Y, \V{\myset{X_k: k\in\comm{}}}\neq Y}  
\end{align}
Since $j\notin \mathcal{C}$, the opinion $X_j$ is conditional independent of $\V{\myset{X_k: k\in\comm{}}}$ and thus, whether an opinion and votes over another set of opinions is correct or incorrect is independent of one another. Formally:
\begin{equation}
	\ind{X_i = Y} \perp \ind{\V{\myset{X_k: k\in\comm{}}}=Y}.
\end{equation}
Therefore, one can write: 
\begin{align}
	\pp{i} &=  \prob{X_i = Y}\prob{\V{\myset{X_k: k\in\comm{}}}=Y} + \prob{X_i \neq Y} \prob{\V{\myset{X_k: k\in\comm{}}}\neq Y},  \\
	&=  p_i \prob{\V{\myset{X_k: k\in\comm{}}}=Y}  + (1-p_i) \yay{1-\prob{\V{\myset{X_k: k\in\comm{}}}=Y}}, \\
	&= p_i p_{\mathcal{C}}  + (1-p_i) \yay{1-p_{\mathcal{C}}}.
\end{align}
Now consider the difference between pseudo competences:
\begin{align}
	\pp{i}-\pp{j} 
	&= p_i p_{\mathcal{C}}  + (1-p_i) \yay{1-p_{\mathcal{C}}} - p_j p_{\mathcal{C}}  + (1-p_j) \yay{1-p_{\mathcal{C}}}, \\
	&=\yay{p_i-p_j}p_{\mathcal{C}} + \yay{1-p_i-\yay{1-p_j}}\yay{1-p_{\mathcal{C}}}, \\
	&=\yay{p_i-p_j}p_{\mathcal{C}} - \yay{p_i-p_j}\yay{1-p_{\mathcal{C}}} \\
	&=\yay{p_i-p_j}\yay{2p_{\mathcal{C}}-1},
\end{align}
which yields that as long as $p_{\mathcal{C}}>\nicefrac{1}{2}$, the pseudo competence preserves the ordering of true competences: $p_i\gtrless p_j \iff \pp{i} \gtrless \pp{j}$.
\section{The Pseudo Competence of Experts that Belong to the Same Committee}
\label{app:morepseudo}
When the pseudo competence of experts that belong to a committee $\mathcal{C}\subset [M]$ is concerned, the definition of pseudo competence is modified to exclude vote of the expert whose competence is to be inferred. Formally, if $i\in\mathcal{C}$, then the pseudo competence $\pp{i}$ is defined as:
\begin{equation}
	\pp{i} \triangleq \prob{X_i = \V{X_k: k\in\comm{}\setminus \myset{i}}}.
\end{equation}
Note that different from the definition in eq. \eqref{pseudocompdef} and the concomitant properties listed in Proposition \ref{prop}, peers of each expert have collectively different competences. Therefore, further refinement for the ordering property is necessary: First, let us define the random variable of correctness for each experts:
\begin{equation}
	\eta_i(t) = \ind{X_i(t) =Y(t)}, \forall t\in [T].
\end{equation} 
Observe that $\eta_i(t)$ is a Bernoulli random variable with parameter $p_i$, $\forall t\in[T]$ and $\eta_i(t) \perp \eta_j(t)$, $\forall i\neq j\in[M]$ at all times. Therefore, given that $i,j\in\mathcal{C}$, the law of total probability yields that we can write the pseudo competence as:
\begin{align}
	\pp{i} &= \sum_{\eta_i\in\myset{\pm 1}}\sum_{\eta_j\in\myset{\pm 1}}\condprob{X_i = \V{X_k: k\in\comm{}\setminus \myset{i}}}{\eta_i,\eta_j} \prob{\eta_i,\eta_j}, \\ 
	&= \sum_{\eta_i\in\myset{\pm 1}}\sum_{\eta_j\in\myset{\pm 1}}\condprob{X_i = \V{X_k: k\in\comm{}\setminus \myset{i}}}{\eta_i,\eta_j} \prob{\eta_i}\prob{\eta_j}.
\end{align}
The last line following from  $\eta_i\perp \eta_j$. Further observe that:
\begin{align}
	&\condprob{X_i = \V{X_k: k\in\comm{}\setminus \myset{i}}}{\eta_i,\eta_j} \prob{\eta_i} \nn \\
	&= \brac{\condprob{\V{X_k: k\in\comm{}\setminus \myset{i}}=Y}{\eta_j} p_i +\condprob{\V{X_k: k\in\comm{}\setminus \myset{i}}\neq Y}{\eta_j}\yay{1-p_i}}.
\end{align}
Thus, one can explicitly write out the pseudo competence as:
\begin{equation}
	\label{pp_exp}
	\begin{aligned}		
		\pp{i} 
		&= \condprob{\V{X_k: k\in\comm{}\setminus \myset{i}}=Y}{X_j= Y} p_i p_j \\ &\hspace{1em}+\condprob{\V{X_k: k\in\comm{}\setminus \myset{i}}\neq Y}{X_j=Y}\yay{1-p_i}p_j \\	
		&\hspace{1em}+\condprob{\V{X_k: k\in\comm{}\setminus \myset{i}}=Y}{X_j\neq Y} p_i \yay{1-p_j} \\
		&\hspace{1em}+\condprob{\V{X_k: k\in\comm{}\setminus \myset{i}}\neq Y}{X_j\neq Y}\yay{1-p_i}\yay{1-p_j}.
	\end{aligned}
\end{equation}
It is possible to derive $\pp{j}$ by swapping $i$ and $j$ in eq. \eqref{pp_exp}. An important set of observation is as follows: 
\begin{align}
	\condprob{\V{X_k: k\in\comm{}\setminus \myset{i}}=Y}{X_j= Y} &= \condprob{\V{X_k: k\in\comm{}\setminus \myset{j}}=Y}{X_i= Y}, \\
	\condprob{\V{X_k: k\in\comm{}\setminus \myset{i}}\neq Y}{X_j\neq Y} &=
	\condprob{\V{X_k: k\in\comm{}\setminus \myset{j}}\neq Y}{X_i\neq Y},
\end{align} 
which follows from the fact that the rest of the committee $\mathcal{C}\setminus\myset{i,j}$ is the same for both sides of the equations. Then, if the following holds: 
\begin{equation*}
	\frac{\pp{i}-\pp{j}}{p_i-p_j} = \condprob{\V{X_k: k\in\comm{}\setminus \myset{j}}= Y\!}{X_i\neq Y}\!-\condprob{\V{X_k: k\in\comm{}\setminus \myset{j}}\neq Y\!}{X_i= Y},
\end{equation*}
then pseudo competences preserves ordering. 

\section{Proof of Lemma \ref{main_lemma}}
\label{app:mainlemma}
Note that with a fixed committee $\comm{}$ and experts with competences $\myset{p_i: i\in[]M}$using agreement statistics as reward in the unsupervised setup is equivalent to having Bernoulli arms with mean $\myset{\pp{i}, \forall i\in[M]}$ in the supervised setup. Then the difference between rewards become:
\begin{equation}
	\tilde{\Delta}_i = \max_{i\in[M]}\pp{i}-\pp{i} = \yay{2p_{\comm{}}-1} \Delta_i
\end{equation}
by eq. \eqref{pushtogether} in Proposition \ref{prop}, yielding the proposed upper-bounds on regret by resorting to \cite[Theorem 2]{audibert2009minimax} for UCB1,\cite[Theorem 6]{audibert2009minimax} for MOSS, and \cite[Theorem 1]{agrawal2013further} for Thompson sampling. For IMED and KL-UCB, observe that by Pinsker's inequality \cite{cover1999elements}:
\begin{equation}
	d\yay{p,q} \geq 2\yay{p-q}^2 
\end{equation}
regret bounds of \cite[Theorem 1]{garivier2011klucb} for KL-UCB, and \cite[Theorem 5]{honda2015imed} for IMED (due to bounded support for reward distribution) yield the proposed bounds. Of course, the elegant analyses of both KL-UCB and IMED yields much sharper bounds but the use of Pinsker's inequality captures the impact of the unsupervised framework on regret.

\pagebreak
\bibliographystyle{IEEE_ECE}
\bibliography{beebib}

\end{document}